\documentclass[aps,prc,preprint,showpacs,superscriptaddress,graphicx]{revtex4}
\usepackage[dvips]{graphicx}

\begin{document}
\def\eps{\epsilon}
\def\la{\Lambda}
\def\si{\Sigma}
\def\sig{{\Sigma^-}}

\noindent
\title{The hadron-quark phase transition in dense matter and neutron 
stars}

\bigskip
\bigskip

\author{G. F. Burgio}
\affiliation{Istituto Nazionale di Fisica Nucleare, Sezione di Catania,
Corso Italia 57, I-95129 Catania, Italy} 
\author{M. Baldo}
\affiliation{Istituto Nazionale di Fisica Nucleare, Sezione di Catania,
Corso Italia 57, I-95129 Catania, Italy} 
\author{P. K. Sahu}
\affiliation{Institute of Physics, Sachivalaya Marg,
Bubhaneswar-751005, India}
\author{H.-J. Schulze}
\affiliation{Istituto Nazionale di Fisica Nucleare, Sezione di Catania,
Corso Italia 57, I-95129 Catania, Italy}

\date{\today}

\bigskip
\bigskip
\bigskip

\begin{abstract}
We study the hadron-quark phase transition in the interior of neutron
stars (NS's). We calculate the equation of state (EOS) of hadronic 
matter using the 
Brueckner-Bethe-Goldstone formalism with realistic two-body and three-body
forces, as well as a relativistic mean field model.  For  quark
matter  we  employ  the  MIT  bag model constraining the bag constant by
using the indications coming from  the recent experimental  results  
obtained  at  the CERN  SPS on  the formation of a quark-gluon plasma. 
We find necessary to introduce a density dependent bag parameter, and the
corresponding consistent thermodynamical formalism.
We calculate the structure of
NS  interiors  with  the EOS comprising both phases, and we find 
that the NS maximum masses fall in a relatively  narrow  interval,  
$1.4\,M_\odot  \leq M_{\rm  max}  \leq  1.7\,M_\odot$.
The precise value of the maximum mass turns out to be only
weakly correlated with the value of the energy density at the 
assumed transition point in nearly symmetric nuclear matter.
\end{abstract}

\bigskip
\bigskip
\bigskip


\maketitle

\section{Introduction}

The  properties of nuclear matter  at high density play a crucial role
for building models of neutron stars \cite{shapiro}. 
The  observed NS's masses are in the range of  
$\approx (1-2)  M_\odot$  (where  $M_\odot$  is  the  mass of the sun, 
$M_\odot =
1.99\times 10^{33}$g), and the radii are of the order of 10 km. 
The characteristics of the core of the NS's influences most strongly
the value of the  maximum mass. The matter inside this core 
possesses densities ranging from a few times $\rho_0$
$(\approx  0.17\;{\rm  fm}^{-3}$,  the normal nuclear matter density) to
one order of magnitude higher. 
Moreover, the equation of state at such high densities is the
main ingredient to determine the structure parameters of NS's, such as 
mass and radius. Therefore, a detailed  knowledge  of  the
EOS  is required for densities $\rho \gg \rho_0$, where a description of
matter only in terms of nucleons and leptons may be inadequate. In fact,
at densities $\rho \gg \rho_0$ several species of other particles,  such
as  hyperons  and  $\Delta$ isobars, may appear, and meson condensations
may take place; also, ultimately, at very high densities, nuclear matter
is expected to undergo a phase transition to a quark-gluon 
plasma \cite{quark}.
The specific goal of the theory is to study the nature of this plasma
and understand the phase transitions between different states.
However, the exact value of the transition density to  quark  matter  is
unknown and still a matter of recent debate
not only in astrophysics, but also within the theory of high energy heavy ion
collisions.

In this paper, we propose a method to determine a range of values of
the maximum mass of NS's
taking  into  account the phase transition from hadronic matter to quark
matter inside the neutron star. The transition point is constrained 
from recent heavy-ion collision data.
Therefore to perform such calculations, we describe the hadron
phase of matter by using two different equations of state, 
{\rm i.e.}, a microscopic non-relativistic EOS obtained in the
Brueckner-Bethe-Goldstone (BBG) theory \cite{book},  
and  a more phenomenological relativistic mean field model \cite{serot}. 
The deconfined quark matter phase is treated by adopting
the popular MIT bag model \cite{chodos}.
In a previous paper \cite{bbsss} the  
bag ``constant'', $B$, which  is  a  parameter  of  the  bag  model, was
constrained  to  be  compatible  with  the  recent  experimental results
obtained at CERN on the formation of a quark-gluon plasma \cite{heinz},
recently confirmed by RHIC preliminary results \cite{blaizot}.

However, it is not obvious if the information on 
the nuclear EOS from high energy heavy ion
collisions can be related to the physics of neutron star interiors. The 
possible quark-gluon plasma produced in heavy ion collisions is expected to
be characterized by small baryon density and high temperature, while the 
possible quark phase in neutron stars appears at high baryon density and
low temperature. If one adopts for the hadronic phase
a non-interacting gas model of nucleons, antinucleons, and
pions, the original MIT bag model  predicts that the 
deconfined phase occurs at an almost constant value of the
quark-gluon energy density, {irrespective} of the thermodynamical
conditions of the system \cite{gavai}. For this reason, it is
popular to draw the transition line between the hadronic and the quark phase 
at a constant value of the energy density, which was estimated to 
fall in the interval between 0.5 and 2 GeV fm$^{-3}$ \cite{mul}. 
This is consistent with the value of about 1 GeV fm$^{-3}$ reported 
by CERN experiments. The close relation between the physics of
neutron stars and of heavy ion collisions is also emphasized by
a recent conjecture that there could be
three phases in heavy-ion collisions at SPS and RHIC energies,
equivalent to the pure quark phase, mixed phase, and pure hadron phase
appearing in neutron stars.
These three phases correspond to (a) an explosive hard quark-gluon phase,
(b) a mixed soft phase (a sort of plateau), and (c) a hadronic phase. 
Considering these three 
phases in a heavy-ion collision model, the first available RHIC data could
be well described \cite{shuryak}. 

The value of 1 GeV fm$^{-3}$ must be considered only an indicative estimate 
of the
transition energy density at zero or nearly zero temperature, as needed
in neutron star studies, and it appears mandatory to explore the
sensitivity of the results on the precise value of the assumed
transition energy density. In this work we present systematic
calculations of neutron star structure, where the hadronic EOS,
which can be considered well established, is implemented with the possible
transition to the deconfined phase described by different
parametrizations of the MIT bag model. The transition energy density
in nearly symmetric nuclear matter at zero temperature is
allowed to vary within a range of values which can be considered still
compatible with the CERN and RHIC data. The calculations will indicate the
sensitivity of the results on the assumed transition point and the
possible correlation between neutron star properties and the
transition energy density value. In particular, we will see that the maximum
neutron star mass is only weakly correlated with the transition energy
density value.

This paper is organized as follows. In Sec.~II we discuss the EOS for the 
hadronic phase of a neutron star, {\rm i.e.}, the BBG and the relativistic 
mean field models. In Sec.~III we apply the MIT bag model to the
description of the quark phase of the neutron star. In Sec.~IV we
present our results and finally in Sec.~V we draw some conclusions.

\section{Hadronic Phase}

\subsection{Brueckner-Hartree-Fock theory}

The Brueckner--Bethe--Goldstone (BBG) theory is based on a linked cluster 
expansion of the energy per nucleon of nuclear matter (see Ref.~\cite{book},
chapter 1 and references therein).  
The basic ingredient in this many--body approach is the Brueckner reaction 
matrix $G$, which is the solution of the  Bethe--Goldstone equation 

\begin{equation}
G[n;\omega] = v  + \sum_{k_a k_b} v {{|k_a k_b\rangle  Q  \langle k_a k_b|}
  \over {\omega - e(k_a) - e(k_b) }} G[n;\omega], 
\end{equation}                                                           
\noindent
where $v$ is the bare nucleon-nucleon (NN) interaction, $n$ is the nucleon 
number density, and $\omega$ the  starting energy.  
The single-particle energy $e(k)$ (assuming $\hbar$=1 here and throughout 
the paper),
\begin{equation}
e(k) = e(k;n) = {{k^2}\over {2m}} + U(k;n),
\label{e:en}
\end{equation}
\noindent
and the Pauli operator $Q$ determine the propagation of intermediate 
baryon pairs. The Brueckner--Hartree--Fock (BHF) approximation for the 
single-particle potential
$U(k;n)$  using the  {\it continuous choice} is
\begin{equation}
U(k;n) = {\rm Re} \sum _{k'\leq k_F} \langle k k'|G[n; e(k)+e(k')]|k k'\rangle_a,
\end{equation}
\noindent
where the subscript ``{\it a}'' indicates antisymmetrization of the 
matrix element.  
Due to the occurrence of $U(k)$ in Eq.~(\ref{e:en}), they constitute 
a coupled system that has to be solved in a self-consistent manner
for several Fermi momenta of the particles involved. 
In the BHF approximation the energy per nucleon is
\begin{equation}
{E \over{A}}  =  
          {{3}\over{5}}{{k_F^2}\over {2m}}  + {{1}\over{2n}}  
~ \sum_{k,k'\leq k_F} \langle k k'|G[n; e(k)+e(k')]|k k'\rangle_a. 
\end{equation}
\noindent
In this scheme, the only input quantity we need is the bare NN interaction
$v$ in the Bethe-Goldstone equation (1). In this sense the BBG 
approach can be considered as a microscopic one. The nuclear EOS can be 
calculated with good accuracy in the Brueckner two hole-line 
approximation with the continuous choice for the single-particle
potential, since the results in this scheme are quite close to the 
calculations which include also the three hole-line
contribution \cite{song}. 
In  the  calculations reported here, we have used the Paris potential 
\cite{lac80} as the two-nucleon interaction.

However, it is commonly known that  
nonrelativistic calculations, based on purely two-body interactions, fail 
to reproduce the correct saturation point of symmetric nuclear matter,
and three-body forces (TBF's) among nucleons are needed to correct this
deficiency. In this work the so-called Urbana model will be used, 
which consists of an attractive term due to two-pion exchange
with excitation of an intermediate $\Delta$ resonance, and a repulsive 
phenomenological central term \cite{schi}. We introduced the same 
Urbana three-nucleon
model within the BHF approach (for more details see Ref.~\cite{bbb}).
In our approach the TBF is reduced to a density dependent two-body force by
averaging over the position of the third particle, assuming that the
probability of having two particles at a given distance is reduced 
according to the two-body correlation function. 
The corresponding nuclear matter EOS fulfills several 
requirements, namely (i) it reproduces the correct nuclear matter 
saturation point,
(ii) the incompressibility  is compatible
with the values extracted from phenomenology, 
(iii) the symmetry energy is compatible with nuclear phenomenology, 
(iv) the causality condition is always fulfilled.  
The equation of state is displayed in Fig.~1, panel a), for symmetric matter
(solid line) and pure neutron matter (dashed line).

Recently, we have included the hyperon degrees of freedom
within the same approximation to calculate the nuclear EOS needed
to describe the NS interior \cite{bbs}. We have included 
the $\Sigma^-$ and $\Lambda$
hyperons. To this purpose, one needs also
nucleon-hyperon (NY) and hyperon-hyperon (YY) interactions \cite{bbs,barc}.
However, because of a lack of experimental data, the hyperon-hyperon 
interaction has been neglected
in the first approximation in this work, whereas for the NY interaction 
the Nijmegen soft-core model \cite{mae89} has been adopted.
 
Once hyperons and leptons are introduced,
the total EOS can be calculated for a given composition
of the baryon components. This allows the determination of the chemical 
potentials of all species which are the fundamental input for the 
equations of chemical equilibrium:
\begin{eqnarray}
  \mu_n &=& \mu_p + \mu_e \:,
\\
  \mu_e &=& \mu_\mu \:,
\\
  2\mu_n &=& \mu_p + \mu_\Sigma \:,
\\
  \mu_n &=& \mu_\Lambda \:.
\end{eqnarray}
Since we are looking at neutron stars 
after neutrinos have escaped, we set the neutrino chemical potentials 
equal to zero. 
The above equations must be supplemented with two other conditions,
{\rm i.e.}, charge neutrality and baryon number conservation. These are 
\begin{eqnarray}
   \rho_p  &=& \rho_e + \rho_\mu + \rho_\Sigma \:,
\label{e:charge}
\\
   \rho &=& \rho_n + \rho_p + \rho_\Sigma + \rho_\Lambda \:.
\label{e:baryon}
\end{eqnarray}
The last two conditions allow the unique solution of a closed system of 
equations, yielding the equilibrium
fractions of the baryon and lepton species for each fixed baryon density.
The latter determine the actual
detailed composition of the dense matter and
therefore the EOS to be used in the interior of neutron stars.
Finally, from the knowledge of the equilibrium composition one determines
the equation of state, i.e., the relation between 
pressure $P$ and energy density $\epsilon$ as a function of baryon 
density $\rho$. 
It can be easily obtained from the thermodynamical relation 
\begin{equation}
  P = -\frac{dE}{dV}  \:, 
\label{e:press}
\end{equation}
being $E$ the total energy and $V$ the total volume.
Equation (\ref{e:press}) can be explicitly worked out in terms of the 
baryonic and leptonic energy densities $\epsilon_B$ 
and $\epsilon_L$,
\begin{eqnarray}
  P &=& -\frac{dE}{dV}  = P_B + P_L \:,
\\
  P_B &=& \rho^2 \frac{d(\epsilon_B/\rho)}{d\rho} \:,
~~ P_L = \rho^2 \frac{d(\epsilon_L/\rho)}{d\rho} \:.
\label{e:pres}
\end{eqnarray}
The total baryonic energy density $\epsilon_B$ is
obtained by adding the energy densities of each species $\epsilon_i$,
which in turn are calculated by taking into account the interaction 
of the species $i$ with the surrounding medium (see Ref.~\cite{bbs} and 
references therein),  
\begin{eqnarray}
\epsilon_B &=& \epsilon_{N} + \epsilon_{\Sigma} +  \epsilon_{\Lambda}. 
\end{eqnarray}
As far as the leptons are concerned, at those high
densities electrons are a free ultrarelativistic gas,
whereas muons are relativistic. Therefore their energy densities 
$\epsilon_L$
and pressures $P_L$ are well known from textbooks, see {\rm e.g.}, 
Ref.~\cite{shapiro}.

\subsection{Relativistic mean field model}

In the present work, we have also considered a hadronic EOS
based on the framework of a relativistic approach, where
one usually starts from a local, renormalizable field 
theory with baryons and explicit meson degrees of freedom.
The theory is chosen to be renormalizable in order to fix
the coupling constants and the mass parameters by empirical
properties of nuclear matter at saturation. As a starting point,
one chooses the mean field approximation which should be reasonably
good at very high densities (a few times nuclear matter density).

In 1974, Walecka first proposed the mean field model \cite{serot,walecka1}, 
where the coupling
constants were chosen in such a way that it fitted nuclear matter binding
energy and saturation density. However, in this model the value of the nuclear
matter incompressibility at saturation is quite high. In order to reproduce 
the correct value, an extension of the Walecka
model was done later, called the nonlinear Walecka 
model \cite{sahu,wm1,wm2,wm3}. 
This model has been proven quite successful in 
describing the properties of nuclei \cite{serot} over a wide range of 
the periodic table.
Thus it is reasonable to use such a model to describe the hadronic phase in a
region where the nuclear densities are not too large in comparison
with the nuclear matter density.

The equation of state for hadrons is calculated in the framework of 
mean field theory using the nonlinear Walecka Lagrangian \cite{sahu,wm1,wm3}
\begin{widetext}
\begin{eqnarray}
{\cal L}(x) & = & \sum_i {\overline \psi_i} (i\gamma^\mu \partial_\mu 
- m_i + g_{\sigma i}\sigma  
+ g_{\omega i} \omega_\mu \gamma^\mu -  g_{\rho i} {\rho_\mu^a} 
\gamma^\mu T_a ) \psi_i \nonumber \\
& - & \frac{1}{4}\omega^{\mu\nu} \omega_{\mu\nu} + \frac{1}{2} {m_\omega^2} 
\omega_{\mu} \omega^{\mu}
+  \frac{1}{2} (\partial_\mu \sigma \partial^\mu \sigma - {m_\sigma^2} 
\sigma^2) -  \frac{1}{4} {\rho_{\mu\nu}^a} {\rho_a^{\mu\nu}}  \\
& + & \frac{1}{2} {m_\rho^2} {\rho_\mu^a} {\rho_a^{\mu}} 
-  \frac{1}{3} b{m_N} (g_{\sigma N} \sigma)^3 
- \frac{1}{4} c (g_{\sigma N} \sigma)^4 + \sum_l {\overline \psi_l}
(i \gamma^\mu \partial_\mu - m_l) \psi_l. \nonumber
\end{eqnarray}
\end{widetext}
This Lagrangian includes nucleons, $\Lambda$ and $\Sigma^-$ hyperons
(denoted by a subscript $i$), leptons (denoted by $l$), and $\sigma$, 
$\omega$, and $\rho$ mesons. The meson fields interact with baryons through 
linear couplings and the coupling constants are different for non-strange
and strange baryons. The $\omega$ and $\rho$ masses are chosen to be 
their physical masses. 
The equation of state is obtained through the mean field
ansatz. In this case one can define effective masses ($\overline m_i$) and 
chemical potentials ($\overline \mu_i$) for the baryons as
\begin{eqnarray}
{\overline m_i} & = & m_i - g_{\sigma i} \overline \sigma, \nonumber \\
{\overline \mu_i} & = & \mu_i - g_{\omega i} \overline \omega_0 - T_3  
g_{\rho i} \overline \rho_0^3,  
\end{eqnarray}
where $\overline \omega_0$, $\overline \sigma$, and $\overline \rho_0^3$ 
are the 
nonzero vacuum expectation values of the meson fields. $T_3$ is the value of 
the $z$-component of the isospin of baryon $i$, whereas $\mu_i$ is the
bare chemical potential. In neutron star matter the chemical potentials 
must fulfill the conditions of equilibrium under weak interaction,
{\rm i.e.},
\begin{eqnarray}
\overline \mu_n &=& \overline \mu_p + \mu_e \:,~~~~\mu_e = \mu_\mu \:,
\\
  2\overline \mu_n &=& \overline \mu_p + \overline \mu_\Sigma \:,~~~~
\overline  \mu_n = \overline \mu_\Lambda \:.
\end{eqnarray}
Moreover, the condition of charge neutrality (\ref{e:charge})
 must be satisfied as well as the baryon number conservation 
(\ref{e:baryon}).
By minimizing the energy at fixed baryon density, one gets
the mean field values of $\overline \omega_0$, $\overline \sigma$, and 
$\overline \rho_0^3$. Then the expressions of the energy density and pressure 
are readily obtained as
\begin{widetext}
\begin{eqnarray}
\epsilon & = & \frac{1}{2} {m_\omega^2} \overline \omega_0^2
+ \frac{1}{2} {m_\rho^2} (\overline \rho_0^3)^2 
+ \frac{1}{2} {m_\sigma^2} \overline \sigma^2 
+ \frac{1}{3} b m_N (g_{\sigma N} \overline \sigma)^3 \nonumber \\
& + & \frac{1}{4} c (g_{\sigma N} \overline \sigma)^4 
+ \sum_i \epsilon_{\rm FG}(\overline m_i, \overline \mu_i) 
+ \sum_l \epsilon_{\rm FG}(m_l, \mu_l), \\
P & = & \frac{1}{2} {m_\omega^2} \overline \omega_0^2
+ \frac{1}{2} {m_\rho^2} (\overline \rho_0^3)^2 
- \frac{1}{2} {m_\sigma^2} \overline \sigma^2 
- \frac{1}{3} b m_N (g_{\sigma N} \overline \sigma)^3 \nonumber \\
& - & \frac{1}{4} c (g_{\sigma N} \overline \sigma)^4 
+ \sum_i P_{\rm FG}(\overline m_i, \overline \mu_i) 
+ \sum_l P_{\rm FG}(m_l, \mu_l),
\end{eqnarray}
\end{widetext}
where $\epsilon_{\rm FG}$ and $P_{\rm FG}$ represent the noninteracting 
fermion 
contributions to the energy density and the pressure. The nonlinear 
Walecka model has eight parameters out of which five are determined by the
properties of nuclear matter. These are the nucleon couplings to scalar
($g_\sigma/m_\sigma$), isovector ($g_\rho/m_\rho$), and vector mesons
($g_\omega/m_\omega$) and the two coefficients $b$ and $c$. 
These are obtained by fitting saturation values of nuclear matter,
{\rm i.e.}, binding energy per nucleon ($\sim-16$ MeV), baryon density ($\sim$ 
0.15~fm$^{-3}$), and Landau mass
(0.83 $m_N$). The symmetry energy coefficient and the compressibility are
taken equal to respectively $30$ and $260$ MeV, 
the same as in BHF calculations and close to estimated values from monopole
oscillations in nuclei \cite{tole}. 

The other three coupling constant parameters of the hyperon
couplings (ratio of hyperon-meson and nucleon-meson couplings)
are not well known. Since hyperons are not present in 
nuclear matter, these cannot be determined from the nuclear matter 
properties. Moreover, from the analysis of experimental data on hypernuclei,
one cannot fix these parameters in a unique way. Therefore, we fix these
parameters by assuming the potentials experienced by $\sig$ hyperons 
to be the same as 
that of $\Lambda$, e.g., $-30$ MeV in the present calculations. First,
we choose the value of hyperon couplings for scalar mesons as 2/3 (similar 
to the quark counting value for $\Lambda$ and $\sig$). Next, using the above
assumption for the hyperon potentials, the values of the hyperon couplings 
for vector mesons are calculated and the hyperon couplings for isovector 
mesons are set 
equal to those for vector mesons assuming vector dominance. 
For further details, the 
reader is referred to Refs.~\cite{sahu1,sahu2} and references therein.

The resulting equation of state is displayed in Fig.~1, panel b), 
for symmetric matter (solid line) and pure neutron matter (dashed line).

\section{Quark Phase}

We now turn to the description of the bulk properties of uniform quark matter,
deconfined from the $\beta$-stable hadronic matter mentioned in the
previous section, by using the MIT bag model \cite{chodos}.
We begin with the thermodynamic potential of $q$ quarks, where $q=u, d, s$
denote up, down, and strange quarks, expressed as a sum of the kinetic term
and the one-gluon-exchange term \cite{quark,fahri},
\begin{widetext}
\begin{eqnarray}
\Omega_q = & - & \frac{3m_q^4}{8\pi^2} \Big[ 
\frac{\eta_q x_q}{3} (2x_q^2-3) + {\rm ln}(x_q+\eta_q) \Big] \nonumber \\
& + & \frac{3m_q^4\alpha_s}{4\pi^3} \Big\lbrace
2 \Big[ \eta_q x_q - {\rm ln}(x_q+\eta_q) \Big]^2 -
\frac{4}{3} x_q^4 + 2 {\rm ln} (\eta_q) \nonumber \\
& + & 4~{\rm ln}(\frac{\sigma_{\rm ren}}{m_q \eta_q} ) \Big[ 
\eta_q x_q - {\rm ln}(x_q +\eta_q) \Big] \Big\rbrace,
\end{eqnarray}
\end{widetext}
where $m_q$ and $\mu_q$ are the $q$ current quark mass and chemical potential, 
respectively, and $x_q = {\sqrt{\mu_q^2 - m_q^2}}/{m_q}$,
$\eta_q = \sqrt{1 + x_q^2} = \mu_q/m_q$.  
$\alpha_s$ denotes the QCD fine structure 
constant, whereas
$\sigma_{\rm ren}$ is the renormalization point, $\sigma_{\rm ren}$ = 313~MeV.
In this work we will consider massless $u$ and $d$ quarks,
in which case the above expression reduces to 
\begin{equation}
\Omega_q =  -  \frac{\mu_q^4}{4\pi^2} 
\Big( 1 - \frac{2\alpha_s}{\pi} \Big), ~~~~~~~
(q=u, d).
\end{equation}
The number density $\rho_q$ of $q$ quarks is related to $\Omega_q$ via
\begin{equation}
\rho_q = - \frac{\partial\Omega_q}{\partial\mu_q},
\end{equation}
and the total energy density and pressure for the quark system 
are given by
\begin{eqnarray}
\epsilon_Q & = & \sum_q (\Omega_q + \mu_q \rho_q) + B,  \label{e:eosqm}\\
P_Q & = & - \sum_q \Omega_q - B, \label{e:pqm}
\end{eqnarray}
where $B$ is the energy density difference between 
the perturbative vacuum and the true vacuum, {\rm i.e.}, the bag constant.
In the  original  MIT
bag   model   the   bag  constant  has  the  value  $B  \approx  55\,\rm
MeV\,fm^{-3}$,  which  is  quite  small  when  compared  with  the  ones
($\approx  210\,\rm  MeV\,fm^{-3}$)  estimated from lattice calculations
\cite{satz1}. In this sense $B$ can be considered as a free parameter.
The composition of $\beta$-stable quark matter is determined by 
imposing the condition of equilibrium under weak interactions for the
following processes
\begin{eqnarray}
u + e^- & \rightarrow & d + \nu_e, \\
u + e^- & \rightarrow & s + \nu_e, \\
d & \rightarrow & u + e^- + \overline \nu_e, \\
s & \rightarrow & u + e^- + \overline \nu_e, \\
s + u & \rightarrow & d + u. 
\end{eqnarray}
In neutrino-free matter ($\mu_{\nu_e} = \mu_{\overline \nu_e} = 0$ ),
the above equations imply for the chemical potentials
\begin{eqnarray}
\mu_d & = & \mu_s = \mu,  \\
\mu & = & \mu_u + \mu_e.  
\end{eqnarray}
As in baryonic matter, the relations for chemical equilibrium must be 
supplemented with the charge neutrality condition and the total baryon 
number conservation :
\begin{equation}
\frac{2}{3}\rho_u - \frac{1}{3}\rho_d - \frac{1}{3}\rho_s - \rho_e = 0,
\end{equation} 
\begin{equation}
\rho = \frac{1}{3} (\rho_u + \rho_d + \rho_s). 
\end{equation} 
It can easily be demonstrated that, in the case of massless $u$, $d$, and $s$
quarks, the equilibrium solution reads
\begin{equation}
\rho_u = \rho_d = \rho_s, ~~~~ì \rho_e = 0,
\end{equation}
and consequently the equation of state is 
\begin{equation}
P_Q  = \frac{1}{3} (\epsilon_Q - 4B). 
\end{equation}
Here one should notice that the above expressions hold in the case 
of constant $B$. If the bag constant is density dependent, all
thermodynamical relations must be reformulated \cite{peng}. 

In this case it is convenient to
consider  the first two terms on the right hand side of Eq.~(\ref{e:eosqm}) 
as a function of density. 
The density of each flavor component $q$ is related to the Fermi momentum
$p_F^{(q)}$  in the usual way,
\begin{equation}
\rho_{\rm q} = {g\over 6\pi^2} [p_F^{(q)}]^3,
\end{equation}
\noindent
where $g$=6 is the spin and color degeneracy factor. 
If we denote by $\rho$ the total baryon density,
the chemical potential $\mu_q$ for each flavour component $q$ can be written
\begin{equation}
\mu_q = E_F^{(q)} + {d B\over d\rho_q} \label{e:mu_b},
\end{equation}
\noindent
where $E_F^{(q)}$ is the kinetic Fermi energy for the $q$ component,
eventually including the perturbative corrections.
The second term on the right hand side of Eq.~(\ref{e:mu_b})
modifies the usual relationship between chemical potential and Fermi energy 
and is absent for a density-independent parameter $B$. This term in turn
modifies the relationship between chemical potential and density in an
obvious way. These additional terms are essential for the consistency
of the different thermodynamical relationships. In particular,
the usual expressions for the pressure

\begin{equation}
P_Q  = \rho {d\epsilon_Q \over \ d\rho} - \epsilon_Q  =
     \sum_q \mu_q \rho_q - \epsilon_Q
\end{equation}
\noindent
hold true provided the additional term involving the derivative of the bag
parameter is included, which is absent in Eq.~(\ref{e:pqm}), where the
free Fermi gas expression for $\Omega_q$ is assumed.
In our calculations both chemical potentials and pressure have been
calculated including the additional terms, thus fulfilling the
correct thermodynamical relationships. We found that this additional term,
coming from the density dependence of the bag parameter, gives a substantial
contribution. In particular, it reduces strongly the
value of the pressure especially in the mixed phase region, as described in 
the next Section.

\section{Results and discussion}

We try to determine a range of possible values for $B$ by exploiting the
experimental  data  obtained  at the CERN SPS, where several experiments
using high-energy beams of Pb nuclei reported the (indirect) evidence  for
the  formation  of  a  quark-gluon  plasma  \cite{heinz}.  The resulting
picture is the following: during  the  early  stages  of  the  heavy-ion
collision,  a very hot and dense state (fireball) is formed, whose energy
materializes in the form of quarks and gluons strongly interacting  with
each other, exhibiting  features  consistent with expectations from a
plasma of deconfined quarks  and  gluons \cite{satz2}.  Subsequently,  the
``plasma"  cools  down and becomes more dilute up to the point where, at
an energy density of about $1\,\rm GeV\, fm^{-3}$ and  temperature  
$T\approx
170\,\rm  MeV$,  the  quarks and gluons hadronize. The expansion is fast
enough so that no mixed hadron-quark equilibrium phase  is  expected  to
occur, and no weak process can play a role. According to the analysis of
those  experiments,  the  quark-hadron  transition  takes place at about
seven times  normal  nuclear  matter  energy  density  ($\eps_0  \approx
156\,\rm MeV\, fm^{-3}$).

In the MIT bag model, the structure of the QCD phase diagram in the
chemical potential and temperature plane is determined by only one
parameter, $B$, although the phase diagram for the transition from
nuclear matter to quark matter is schematic and not yet completely
understood, particularly in the light of recent investigations on a
color  superconducting  phase  of quark matter \cite{khrisna}. 
In our analysis we assume that the transition to a quark-gluon plasma 
is determined by the value of the energy density alone (for a given 
asymmetry). With  this assumption  and  taking  the  hadron  to  quark matter 
transition energy density from the CERN experiments we estimate in the
following the value of $B$ and its possible density dependence. 

\subsection{Phase transition in symmetric nuclear matter}
First, we  calculate  the  EOS  for  cold  asymmetric  nuclear  matter
characterized by a proton fraction $x_p = 0.4$ (the one  for  Pb  nuclei
accelerated at CERN-SPS energies) in the BHF formalism with two-body and
three-body forces (as described earlier). We perform the same calculation 
in the RMF approach using $x_p = 0.5$. 
Then we calculate the EOS for $u$ and $d$ quark matter using 
Eq.~(\ref{e:eosqm}).
First we use a constant, density independent, $B$. 
The results are shown in  Fig.~2. 
The solid lines represent the BHF (left panel) and the RMF (right panel)
calculation of the energy density. The dotted (dashed) lines represent
the quark matter equation 
of state calculated for $B=55~ {\rm MeV~ fm^{-3}}$ 
($B=90~ {\rm MeV~ fm^{-3}}$), and several values of the QCD coupling 
constant $\alpha_s$. We find that at  very  low  baryon
density  the  quark matter energy density is always higher than that of nuclear
matter, independently of the value of $B$. Therefore nuclear matter is the 
favourite state. However, 
for $B=55~ {\rm MeV~ fm^{-3}}$, the two energy  densities
become equal at a certain value of nuclear density. 
Unfortunately this crossing takes place at normal nuclear matter density,
both in the BHF and the RMF approach. Therefore we try a larger value of $B$.
Since the expression of the quark matter energy density is linear in $B$
[see Eq.~(\ref{e:eosqm})], an increase of $B$ means an overall shift 
towards larger energy densities. The two curves can cross now at a 
slightly larger baryon density, but still much smaller than the   
desired point, {\rm i.e.}, $E/V  \approx  7\eps_0  \approx
1.1\,\rm  GeV\,fm^{-3}$. No crossing at all is present above some limiting 
value of $B$. In Fig.~2 we show for completeness also the limiting case of 
$B=90~ {\rm MeV~ fm^{-3}}$, where the nuclear matter
energy density is always smaller than the one of quark matter.
These results are not very sensitive to the value of $\alpha_s$. 

Therefore, we assume a density dependent $B$ (an eventual dependence of $B$
on the asymmetry $x_p$ is not considered at this stage). In the literature
there  are  attempts  to  understand  the  density  dependence  of   $B$
\cite{brown,blaschke};  however, currently the results are highly 
model dependent
and no definite picture has come  out  yet.  Therefore,  we  attempt  to
provide  effective  parametrizations for this density dependence, trying
to  cover  a  wide  range  by  considering  some  extreme  choices.  Our
parametrizations  are  constructed  in  such  a  way  that at asymptotic
densities $B$ has some finite value $B_\infty$. 
In order to fix $B_\infty$ we proceed in the following way. 
The energy density for $u$, $d$ quark matter reads
\begin{widetext} 
\begin{equation}
\epsilon_Q(\rho,x_p) =  B(\rho)  
+ \frac{3}{4} \Big\lbrack \frac{\pi^2}{
(1 - 2\alpha_s/\pi)} \Big\rbrack ^{1/3} \Big\lbrack 
(1+x_p)^{4/3} + (2-x_p)^{4/3} \Big\rbrack ~\rho^{4/3}.
\label{e:etran}
\end{equation}
\end{widetext}
$B_\infty$ can be readily calculated at the transition energy density 
(known from experiments) which corresponds to a value of the baryonic 
number density $\bar \rho$ given by the hadronic equation of state, 
{\rm i.e.},
\begin{widetext}
\begin{equation}
B_\infty = \epsilon_Q(\bar\rho,x_p)  
 - \frac{3}{4} \Big\lbrack 
\frac{\pi^2}{
(1 - 2\alpha_s/\pi)} \Big\rbrack ^{1/3} \Big\lbrack 
(1+x_p)^{4/3} + (2-x_p)^{4/3} \Big\rbrack ~\bar\rho^{4/3}.
\end{equation}
\end{widetext}
Therefore we can determine a range of values for $B_\infty$, 
that are shown in Table 1.

We limit ourselves to consider only two possible values of $\alpha_s$, 
{\rm i.e.} $\alpha_s=0$ and $\alpha_s=0.1$, here and throughout this paper. 
Although the values of $B_\infty$ span a wide range, 
we have verified that our results do not 
change appreciably by varying this value, since at large densities the quark
matter   EOS   is   dominated   by  the  kinetic  term  on  the  RHS  of
Eq.~(\ref{e:etran}). With those values of $B_\infty$ we then construct
two parametrizations of $B$ as function of the baryon density.
First, we use a Gaussian parametrization given as
\begin{eqnarray}
B(\rho)  =  B_\infty  +  (B_0  -  B_\infty)  \exp  \left[  -\beta  \Big(
\frac{\rho}{\rho_0} \Big)^2 \right] \:. \label{e:g}
\end{eqnarray}
The parameter $\beta$ is fixed numerically by imposing that the 
quark matter energy density from  Eq.~(\ref{e:etran}) matches the nucleonic 
one at the desired transition density $\bar \rho$.
Therefore  $\beta$ depends only  on  the  free  parameter  $B_0 =
B(\rho=0)$. However, the exact value of $B_0$ is not very  relevant  for
our  purpose,  since  at  low  density  the matter is in any case in the
nucleonic phase. In our previous paper \cite{bbsss} we have used 
both $B_0  =  200\,\rm  MeV\,fm^{-3}$ and $B_0  =  400\,\rm  MeV\,fm^{-3}$,
and found that the results did not sensitively depend on it.
Therefore in this work we limit ourselves to use the value
$B_0  =  400\,\rm  MeV\,fm^{-3}$.

We also use another  extreme,  Woods-Saxon like, parametrization,
\begin{eqnarray}
B(\rho)  =  B_\infty + (B_0 - B_\infty) \left[1 + \exp\left(\frac{\rho -
\tilde \rho}{\rho_d} \right)\right]^{-1} \:, \label{e:ws}
\end{eqnarray}
where $B_0$ and $B_\infty$ have the same meaning as described before for
Eq.~(\ref{e:g}) and $\tilde\rho$ has been fixed in the same way as $\beta$
for  the Gaussian parametrization. However, we have chosen sets of values for
$\tilde\rho$ and $\rho_d$ in such a way that $B$
remains practically constant at a value $B_0$ up to  a  certain  density
and  then  drops  to $B_\infty$ almost like a step function.
It  is   an   extreme parametrization  in  the sense that it will delay 
the onset of the quark phase in neutron star matter as much as possible.
The parametrizations of $B$ [Eqs.~(\ref{e:g}) and (\ref{e:ws})]
are shown in Fig.~3 for BHF and Fig.~4 for RMF equations of state.

The complete results for the energy densities are shown in Fig.~5
for the BHF and Fig.~6 for the RMF nucleonic equation of state
(solid lines). By assuming that the hadron-quark transition 
takes place within a range of energy density values, we have considered 
three possible values of the transition energy 
density, {\rm i.e.}, $\epsilon_Q = 0.8, ~1.1$, and 1.5 GeV
$\rm fm^{-3}$. We find that at very low baryon density the quark matter
energy density is higher than that of nuclear matter, while with increasing
baryon density the two energy densities become equal at a certain point
(indicated by the full dot), and after that the nuclear matter energy density 
remains always higher. We identify this crossing point with the
transition density from nuclear matter to quark matter.

\subsection{Phase transition in beta-stable neutron star matter}

With  these  parametrizations  of  the  density dependence of $B$ we now
consider  the  hadron-quark  phase  transition  in  neutron  stars.
In both the BHF and the RMF approach, we calculate the EOS of a 
conventional neutron star as
composed  of  a  chemically  equilibrated  and charge neutral mixture of
nucleons, hyperons, and leptons. The  result  is  shown  by  the  
solid  lines  in Figs.~7 and 8, respectively.

The dotted (dashed) lines represent the EOS of beta-stable and
charge neutral ($u$,$d$,$s$) quark matter obtained within the MIT 
bag model, with $B$ parametrized as a Gaussian-like (Woods-Saxon-like) 
function.
In particular, the left-hand panels display calculations with a 
transition energy density $\epsilon_Q= {\rm 0.8~ GeV~ fm^{-3}}$, 
whereas the central panels
show the results for $\epsilon_Q= {\rm 1.1~ GeV~ fm^{-3}}$ and 
the right-hand panels for $\epsilon_Q= {\rm 1.5~ GeV~ fm^{-3}}$. 
Two sets of values of the $s$-quark mass and the QCD coupling 
constant $\alpha_s$ 
are considered, namely $m_s={\rm 150~MeV},\alpha_s=0$ (upper panels) and
$m_s={\rm 200~MeV},\alpha_s=0.1$ (lower panels). 
The full squares (diamonds) represent the crossing points
between the hadron and the quark phase. They lie inside the mixed 
phase region, whose range, at this stage, cannot be derived from 
the behavior of the energy density alone. In spite of that,
some qualitative considerations can be done. In fact, we notice that
the mixed phase starts at values of the baryon density which increase 
with increasing $\epsilon_Q$,
and that those values turn out to be weakly dependent on the values 
of $m_s$ and $\alpha_s$. However, if $\epsilon_Q= {\rm 1.5~ GeV~ fm^{-3}}$, 
no phase transition at all is present when the Woods-Saxon-like 
parametrization of $B$ is 
adopted and the BHF EOS is considered. In this case, neutron star matter
always remains in the hadronic phase. This does not hold for the RMF EOS 
(see Fig.~8), where a well defined phase transition is present with each
parametrization chosen for $B$. In this case the crossing points
are shifted to values of baryonic densities slightly smaller than in the
BHF case. Therefore, in the RMF case, the onset of the mixed phase should
start earlier.

Now we are ready to perform the Glendenning construction \cite{glen},
which determines the range of baryon density where both
phases coexist. The essential point of
this procedure is that both the hadron and the quark phase are allowed to be 
separately charged, still preserving the total charge neutrality. 
This implies that neutron star matter can be treated as a two-component
system, and therefore can be parametrized by two chemical potentials.
Usually one chooses the pair ($\mu_e, \mu_n$), {\rm i.e.}, electron and 
baryon chemical potential. The pressure is the same in the two 
phases to ensure mechanical stability, while the chemical potentials of
the different species are related to each other satisfying chemical and
beta stability. The Gibbs condition for mechanical and chemical equilibrium 
at zero temperature between both phases reads

\begin{equation}
P_{\rm {HP}}(\mu_e, \mu_n) =P_{\rm{QP}}(\mu_e, \mu_n) = P_{\rm {MP}}. 
\label{e:mp}
\end{equation}
From this equation we can calculate the equilibrium chemical potentials 
of the mixed phase corresponding to the intersection 
of the two surfaces representing the hadron and the quark
phase. At densities below the mixed phase, the system is in the 
charge neutral hadronic phase, and above the pressure of the charge
neutral quark phase is higher than the one in the mixed phase. Therefore 
the system is in the quark phase. The intersection of the two surfaces
allows one to calculate the charge densities $\rho_c^{\rm{HP}}$ and
$\rho_c^{\rm{QP}}$ and therefore the volume fraction $\chi$ occupied 
by quark matter in the mixed phase, {\rm i.e.},
\begin{equation}
\chi \rho_c^{\rm{QP}} + (1 - \chi) \rho_c^{\rm{HP}} = 0.
\label{e:chi}
\end{equation}
From this, the energy density $\epsilon_{\rm{MP}}$ and the baryon density 
$\rho_{\rm{MP}}$ of the mixed phase can be calculated as
\begin{eqnarray}
\epsilon_{\rm{MP}} &=& \chi \epsilon_{\rm{QP}} + (1 - \chi) 
\epsilon_{\rm{HP}}, \\
\rho_{\rm{MP}} &=& \chi \rho_{\rm{QP}} + (1 - \chi) 
\rho_{\rm{HP}}. \label{e:mp1}
\end{eqnarray}
The resulting EOS's for neutron star matter, according to
the different bag parametrizations and hadronic EOS's, are reported in 
Figs.~9, 10, and 11 for  $\epsilon_Q={\rm 0.8,~ 1.1,~ and~ 1.5~GeV~ fm^{-3}}$,
respectively. In those figures, we display the pressure vs.~the baryon 
density for the chosen parametrizations of $B$. The left (right)-hand panels 
represent the calculations performed with the BHF (RMF) EOS.
The shaded area represents the mixed phase.
A pure quark phase is present at  densities
above the shaded area  and a pure hadronic phase is present below it.

We note that, for a low transition density 
$\epsilon_Q ={\rm 0.8~GeV~fm^{-3}}$ (see Fig.~9), 
the pure hadron phase can be completely absent in some cases 
(see panels a,b,c,e), 
whereas in the other cases (panels d,f,g,h) a small hadronic component 
is always present. The mixed phase starts at low baryon densities, well
below the threshold for hyperon formation, and extends up to 0.65 (0.77)
${\rm fm^{-3}}$ when the bag constant is parametrized with a 
Gaussian-like (Woods-Saxon like) function. This result holds for both 
the BHF and RMF description of the hadronic phase.
Therefore, when the quark-hadron transition takes place at
$\epsilon_Q = {\rm 0.8~GeV~fm^{-3}}$ in symmetric matter, the corresponding
neutron stars are characterized in some cases by the absence of a crust,
with a mantle made of a mixed phase plus a pure quark phase in the core. 
In other cases, neutrons stars will have a crust and a very thin, purely
hadronic layer, followed by a large mixed phase and a heavy quark core.

When the transition in symmetric matter takes place at 
$\epsilon_Q = {\rm 1.1 ~GeV~fm^{-3}}$ (Fig.~10) \cite{bbsss}, we observe
naturally a shift of the onset of the mixed phase towards larger 
baryonic densities, although it turns out to be still slightly
smaller than the density for hyperon formation in pure hadronic
matter. Of course hyperons are still present in the hadron component
of the mixed phase. In particular, when the Gaussian-like (Woods-Saxon-like)
parametrization of the bag constant $B$ is used, the mixed phase extends 
from 0.15--0.2 (0.21--0.25) ${\rm fm^{-3}}$ up to 0.7 (0.9) 
${\rm fm^{-3}}$, with a 
slight dependence on the $s$-quark mass and the QCD coupling constant 
$\alpha_s$. In all cases, from Fig.~10 we notice that the hadronic phase 
is always present, although it is limited to a narrow range of low densities.
In this case, neutron stars will always possess a hadronic layer and a crust.
 
Finally, when the transition in symmetric matter takes place at 
$\epsilon_Q = {\rm 1.5 ~GeV~fm^{-3}}$ (Fig.~11), a new scenario can show up.
In fact, when the Woods-Saxon-like parametrization is used for $B$ and
the BHF EOS is used for the hadronic component (panels c, d), no phase 
transition to quark matter is observed, and neutron star matter remains in 
the hadronic phase. Pure hyperon stars are then produced. In the RMF case
(panels g, h) the onset of the mixed phase is shifted to 0.4--0.47
${\rm fm^{-3}}$ and extends up to about 1.25 ${\rm fm^{-3}}$. 
This gives rise to neutron stars with a thick hadronic layer. 
However, when the Gaussian-like parametrization is used, the mixed phase 
starts at the same low densities as before and extends up to 0.84 (0.77)
${\rm fm^{-3}}$ when the BHF (RMF) EOS is adopted.

Therefore, when the Gaussian-like parametrization of $B$ is used, 
the onset of the mixed phase is localized at low densities and
remains almost constant with changing the transition energy density
in symmetric matter. If a Woods-Saxon-like parametrization is chosen, the
mixed phase will start at higher baryon densities, its value 
changing according to the value of the transition energy density in 
symmetric matter. In this way, we are exploring a whole set of possible 
scenarios of quark-hadron phase transitions, and a corresponding set of 
neutron star configurations. 

It has to be stressed that when the Gaussian parametrization for $B$
is used, the density of the hadron component reaches only moderate
high values. The highest value is obtained at the end of the mixed phase,
where pure quark matter appears. When the transition density is fixed
at $\epsilon_Q = $ 1.1 GeV fm$^{-3}$, this maximum hadron density is about
2.5 times the saturation density, with no hyperon component. For
such a range of density the hadron EOS can be considered well established,
with only little uncertainities. When $\epsilon_Q = $ 1.5 GeV fm$^{-3}$
is assumed, the maximum hadron density ia about 4 times saturation
density, but with a 20 \% content of hyperons. For these density values
the microscopic theories can still produce reliable predictions on
the hadron EOS. Similar considerations apply when the Wood-Saxon
parametrization is considered, except of course when no mixed phase
is present. This parametrization, however, has to be considered
less realistic, since it is devised to shift to artificially high density
the onset of the quark phase.

\subsection{Structure of neutron stars}

We assume that a neutron star is a spherically symmetric distribution of 
mass in
hydrostatic equilibrium. The equilibrium configurations are obtained
by solving the Tolman-Oppenheimer-Volkoff (TOV) equations \cite{shapiro} for 
the pressure $P$ and the enclosed mass $m$,
\begin{widetext}
\begin{eqnarray}
  {dP(r)\over{dr}} &=& -{ G m(r) \epsilon(r) \over r^2 } \,
  {  \left[ 1 + {P(r) / \epsilon(r)} \right] 
  \left[ 1 + {4\pi r^3 P(r) / m(r)} \right] 
  \over
  1 - {2G m(r)/ r} } \:,
\\
  {dm(r) \over dr} &=& 4 \pi r^2 \epsilon(r) \:,
\end{eqnarray}
\end{widetext}
being $G$ the gravitational constant. 
Starting with a central mass density $\epsilon(r=0) \equiv \epsilon_c$,  
we integrate out until the pressure on the surface equals the one 
corresponding to the density of iron.
This gives the stellar radius $R$ and the gravitational mass is then 
\begin{equation}
M_G~ \equiv ~ m(R)  = 4\pi \int_0^Rdr~ r^2 \epsilon(r) \:. 
\end{equation}
We have used as input the equations of state displayed in Figs.~9, 10, 
and 11. For the description of the NS's crust, if present, we have joined 
the hadronic 
equations of state with the ones by Negele and Vautherin \cite{nv}
in the medium-density regime, and the ones   
by Feynman-Metropolis-Teller \cite{fey} and Baym-Pethick-Sutherland 
\cite{baym} for the outer crust. 

The results are plotted in Figs.~12, 13, and 14.
We display the gravitational mass $M_G$ (in units of the solar mass $M_\odot$)
as a function of the radius $R$ (left-hand panels) and central baryon density 
$\rho_c$ (right-handed panels). The solid line represents the calculation
for beta-stable asymmetric nuclear matter including hyperons.
We note that the inclusion of hyperons gives a low
value of the maximum mass equal to 1.26 $M_\odot$ in the BHF case.
This value lies below the best 
observed pulsar mass, PSR1916+13, which amounts to 1.44 solar 
masses \cite{hulse}.
In the case of the RMF model, the corresponding EOS produces values of 
the maximum mass close to $1.7\,M_\odot$.    

The possible occurrence of a quark core is usually assumed to
further soften the EOS and lower the maximum mass. This is indeed the case 
in the RMF model, as apparent in Figs.~12, 13, and 14 (lower panels). 
However, the situation is reversed
in the BHF case, where the  EOS  becomes, on the contrary, stiffer.
Correspondingly, the inclusion of the quark component has the
effect of increasing the maximum mass in the BHF case and of decreasing it
in the RMF case. This can be clearly seen in 
Figs.~12, 13, and 14, which display the results with a transition 
energy density in symmetric matter equal to 0.8, 1.1, and 1.5 
${\rm {GeV~fm^{-3}}}$, respectively.
The maximum value of the neutron star mass lies in the range
$ 1.44 M_\odot \leq M_{\rm max} \leq 1.7 M_\odot$, 
independent of the EOS used for the hadronic component, and no matter 
which parametrization chosen for $B$. It depends only weakly on the transition 
energy density in symmetric matter.  
The configurations of the NS with a quark core are characterized by 
a smaller radius and a higher value of the central density, compared 
to the pure hadronic case (solid lines). In some cases, these stars have no 
crust at all, since in the EOS the hadronic component is missing.  

In Fig.~15 we plot a typical density profile for a star with canonical
mass 1.4 $M_\odot$, obtained when the transition energy
density in symmetric matter is equal to 1.1 $\rm GeV~fm^{-3}$ \cite{bbsss} 
and $B$ has been parametrized as a Woods-Saxon-like function, with two
different choices of $m_s$ and $\alpha_s$. 
On the left-hand panels we plot the result 
obtained when the BHF EOS has been used for the hadronic component, whereas
on the right-hand panels the corresponding case obtained with the
RMF EOS is shown.
We observe that a large part of the core is composed of pure quark 
matter (about 6 km), then a  thick layer of a couple of kilometers
is in the mixed phase, followed by a modest hadronic zone and a thin crust.
This generic profile turns out to be only slightly dependent on the 
EOS used for the hadronic component.  

Some remarks should be done about the behavior of the mixed phase.
As one can see clearly from Figs.~12, 13, and 14, the presence of a mixed 
phase produces
a kind of plateau in the mass vs.~central density relationship,
which is a direct consequence of the smaller slope displayed by
all EOS's in the mixed phase region, see Figs.~9, 10, and 11. In this region,
however, the pressure is still increasing monotonically, 
despite the apparent smooth behavior, and no unstable configuration can 
actually appear. We found that the appearance of this slow variation
of the pressure is due to the density dependence of the bag constant,
in particular the occurrence of the density derivative of the
bag constant in the pressure and chemical potentials, as required
by thermodynamic consistency. To illustrate this point
we calculate the EOS for quark matter with a density independent
value of $B = 90$ MeV fm$^{-3}$, see Fig.~16, and the corresponding
neutron star masses. The EOS is now quite smooth and the mass
vs.~central density shows no indication of a plateau. 

Finally, it has to be pointed out that the maximum mass value, whether $B$ is
density dependent or not, is dominated by the quark EOS at densities where
the bag constant is much smaller than the quark kinetic energy.
The constraint coming from heavy ion reactions, as discussed
above, is relevant only to the extent that it restricts 
$B$ at high density within a range of values, which are commonly
used in the literature. This can be seen also from Fig.~16, where 
the (density independent) value of $B = {\rm 90~MeV}$ produces again
a maximum value around $1.5$ solar masses.  
\par

\section{Conclusions}

We studied neutron star properties, in particular NS's maximum masses,
using  an  EOS  which
combines reliable EOS's for hadronic matter and a bag model EOS for
quark matter.  We found that a density dependent $B$ is
necessary to get the transition to the quark-gluon plasma in nearly
symmetric nuclear matter at an energy
density which is well above saturation density and in a range of values
which can be considered compatible with the CERN-SPS and RHIC findings
on  the  phase  transition from  hadronic  matter  to  quark matter.
We considered a wide range of values, from 0.8 $\rm GeV~fm^{-3}$ to
1.5 $\rm GeV~fm^{-3}$, in order to establish the sensitivity of the results
on the assumed value of the transition energy density.
\par
For a given value of the transition density for symmetric nuclear matter,
the corresponding transition in neutron star matter, i.e., beta
stable matter, occurs in general at substantially lower energy
density. It is essential, in this respect, that in the calculations
strange matter is
included and allowed to develop inside neutron star matter, since
the appearence of strange matter tends in general to soften the EOS.
The results show that the NS maximum mass is clearly correlated with the
assumed value of the transition energy density. For a given
transition density, the maximum mass falls in a narrow range, nearly
independent of the details of the parametrization of the bag model.
As the transition density is made to vary, the value of the maximum mass
is shifted. In general it decreases at increasing value of the transition
energy density if the hadron EOS is computed within the microscopic
BHF scheme. The trend is reversed with the hadron EOS computed within the
relativistic mean field method. However, this correlation appears to be
rather weak, and the full range of possible values of the maximum
mass turns out to be  between 1.4 and 1.7 solar masses. 

The value of the maximum mass is mainly determined by the quark component
of the neutron star and by the corresponding EOS. In this sense,
one can say that the value of the neutron star maximum mass
can be a good testing ground for the quark EOS, rather than the
hadron EOS.
Indeed, the value of the maximum mass of neutron stars obtained according to
our analysis appears robust with respect to the uncertainties 
of the nuclear EOS, and the obtained range of values is mainly due
to the uncertainties of the quark EOS.\par
Other recent calculations of neutron star properties employing
various RMF nuclear EOS's together with either effective mass bag  
model  \cite{bag}  or 
Nambu-Jona-Lasinio  model  \cite{njl}  EOS's  for quark matter, also give
maximum masses of only about $1.7\,M_\odot$, even though not constrained
by hints coming from the CERN-SPS and RHIC data. 
Therefore, according to our results, the experimental observation of a heavy
($M > 1.8 M_\odot$) neutron star, as claimed recently by some groups
\cite{kaaret}($M \approx 2.2 M_\odot$),
if confirmed, would suggest that  
serious drawbacks are present for the possible description 
of the high-density phase of quark matter within the bag model.


\newpage
\begin{table}
\caption{
The values of $B_\infty$ (in MeV fm$^{-3}$) are 
displayed vs. the ones of the energy density (in GeV fm$^{-3}$) 
in $u$, $d$ quark matter at 
the transition point for $\alpha_s=0, 0.1$.  The corresponding values of 
the baryonic densities $\bar\rho$ (in fm$^{-3}$) at transition are 
deduced from the BHF and RMF hadronic equations of state.} 
\bigskip
\begin{ruledtabular}
\begin{tabular}{|ccccc|}
EOS &   $\epsilon_Q$ &  $\bar\rho$ & $B_\infty^{\alpha_s=0}$ & 
$B_\infty^{\alpha_s=0.1}$ \\  
\hline 
BHF  &  0.8 &  0.76 & 36.4 & 19.6\\
&   1.1 & 0.97 & 51.1 &  27.9\\
&   1.5 & 1.22 & 77.4 & 45.8\\
\hline
RMF &  0.8 & 0.76 & 37.9 & 21.0\\
&  1.1 &  0.98 &  37.8 &  14.3\\
&  1.5 &  1.23 & 55.4 & 23.4\\
\end{tabular}
\end{ruledtabular}
\label{t:t1} 
\end{table}

\newpage
.

\begin{figure}
\includegraphics{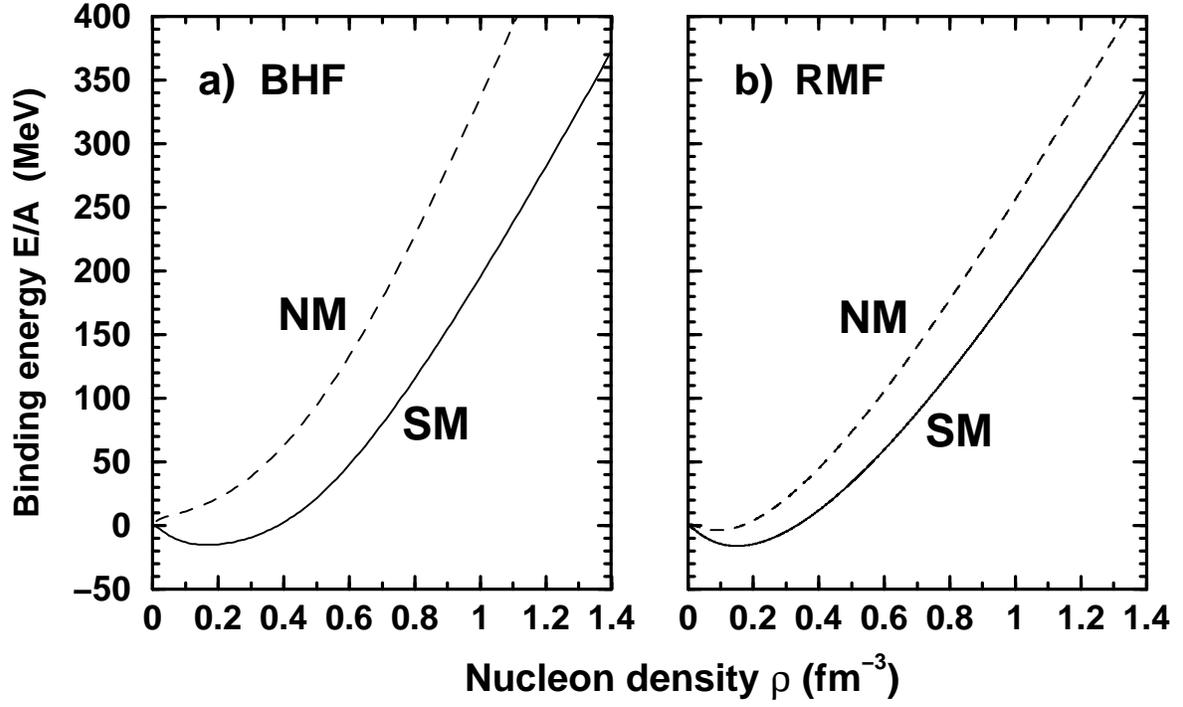}
\caption{
The BHF [RMF] equation of state is displayed in panel a) [panel b)]
for symmetric matter (solid line) and pure neutron matter (dashed line). 
\label{f:f1}}
\end{figure}

\begin{figure}
\begin{center}
\includegraphics{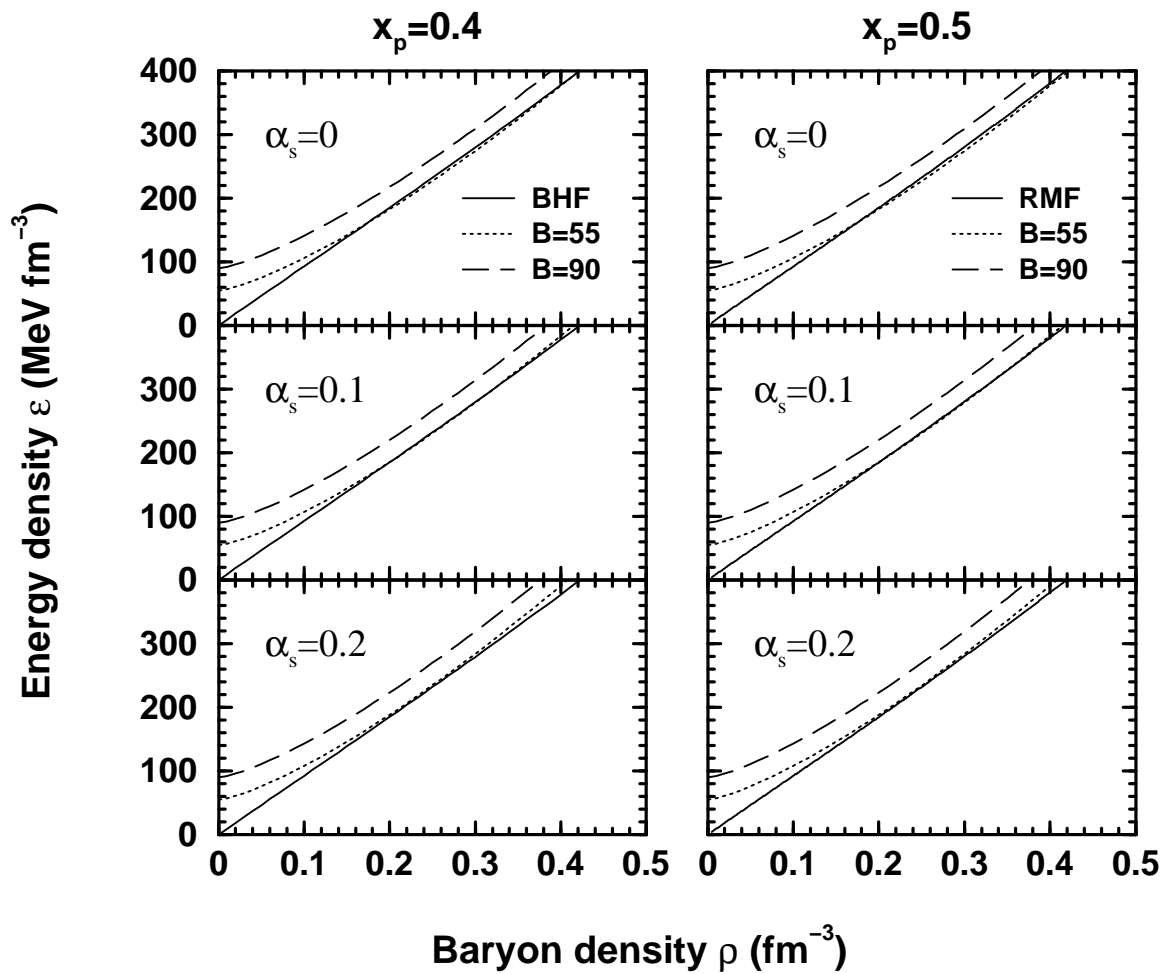}
\end{center}
\caption{The energy density is displayed vs.~the baryon density for 
almost symmetric matter. Panels on the left-hand side show BHF 
calculations ($x_p=0.4$, solid line), whereas panels on 
the right-hand side show the ones obtained with the relativistic mean field 
model ($x_p=0.5$, solid line).
The dotted (dashed) lines represent calculations for $u$,$d$ quark matter 
within the MIT bag model, performed with 
$B=55~{\rm MeV~fm^{-3}}$ ($B=90~{\rm MeV~fm^{-3}}$)
and several values of the QCD coupling constant $\alpha_s$. 
}\label{f:f2}
\end{figure}

\begin{figure}
\includegraphics[width=12cm]{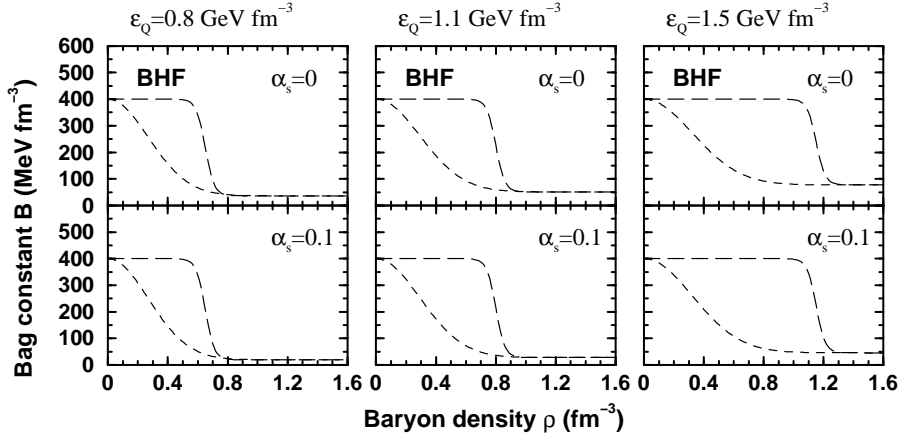}
\caption{The bag constant $B$ is displayed vs.~the baryon density $\rho$.
Two parametrizations are adopted, {\rm i.e.}, a gaussian one 
(short dashed lines) and a Woods-Saxon-like one (long dashed lines).
The left, central, and right panels correspond to different transition
energy densities $\epsilon_Q = {\rm 0.8,~ 1.1,~ and~ 1.5~ GeV~ fm^{-3}}$.
Calculations are performed for the BHF nucleonic equation of state, and
for two values of $\alpha_s=0, 0.1$.
}\label{f:f3}
\end{figure}

\begin{figure}
\includegraphics[width=12cm]{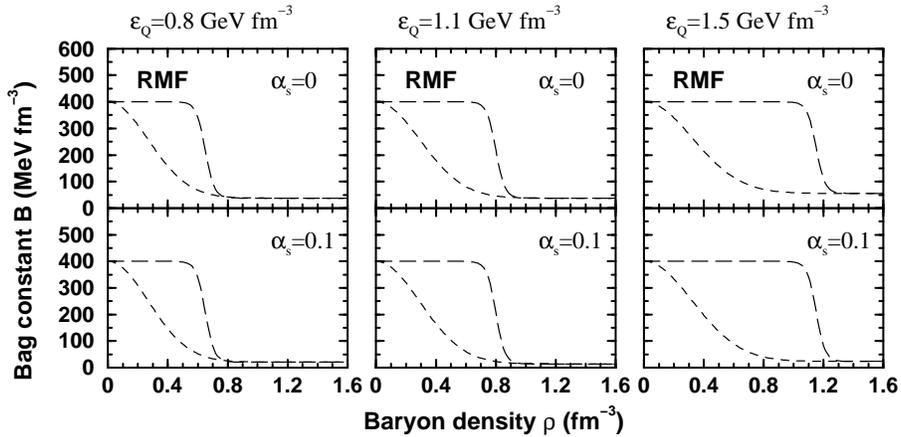}
\caption{Same as Figure 3, but for the RMF nucleonic equation of state.
}\label{f:f4}
\end{figure}

\begin{figure}
\includegraphics[width=13cm]{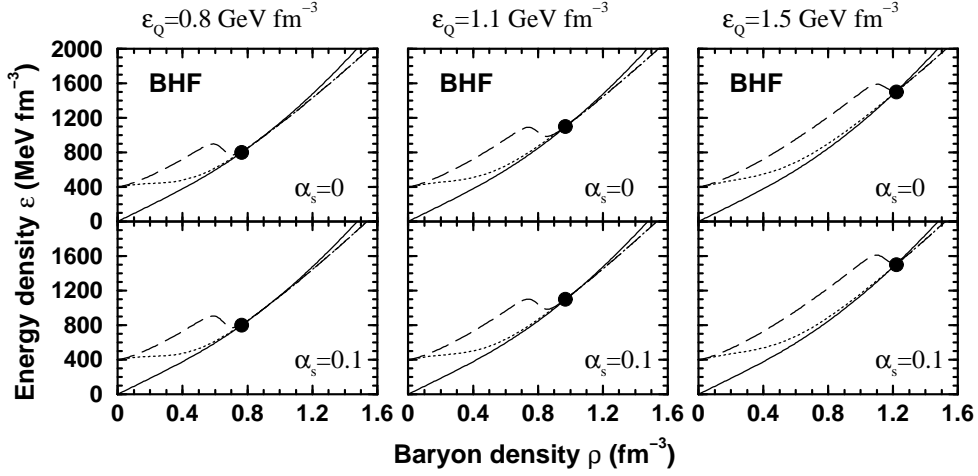}
\caption{The energy density is displayed vs.~the baryon density for 
almost symmetric matter ($x_p=0.4$, solid line) in the BHF approach.
The left, central, and right panels correspond to different transition
energy densities $\epsilon_Q = {\rm 0.8,~ 1.1,~ and~ 1.5~ GeV~ fm^{-3}}$.
Those values are represented by the full dots.
The short (long) dashed lines represent 
calculations for $u$,$d$ quark matter obtained within the MIT bag model, 
with $B$ parametrized as a Gaussian (Woods-Saxon) function.
Two values of the QCD coupling constant $\alpha_s=0, ~0.1$ are considered. 
}\label{f:f5}
\end{figure}

\begin{figure}
\includegraphics[width=13cm]{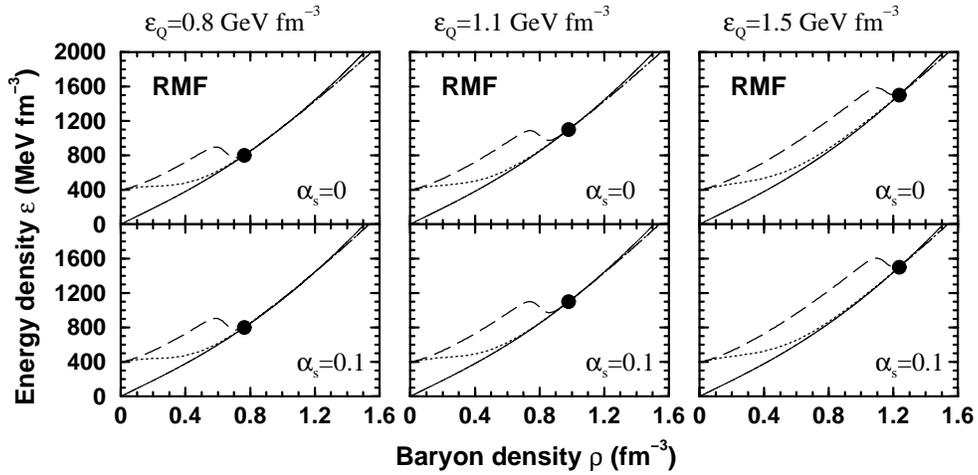}
\caption{Same as Figure 5, but for the RMF hadronic equation of state with
$x_p=0.5$.
}\label{f:f6}
\end{figure}

\begin{figure}
\includegraphics[width=13cm]{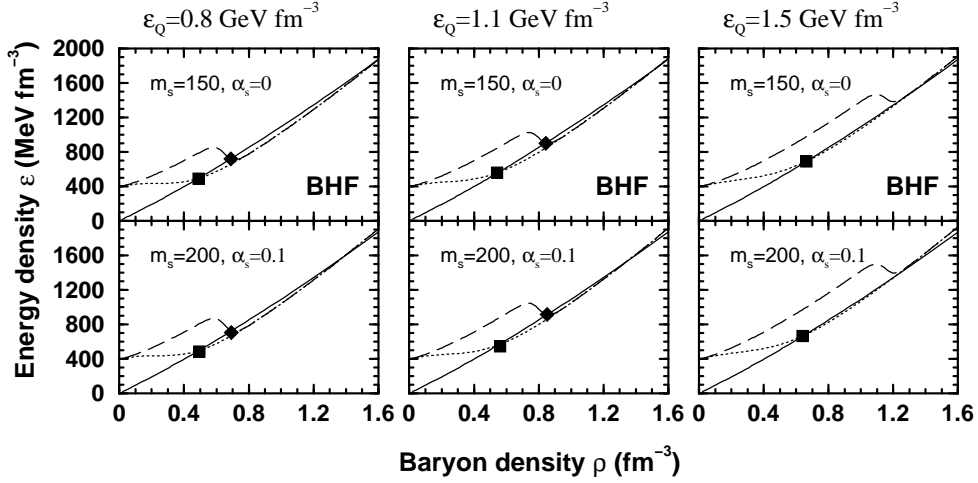}
\caption{The energy density is displayed vs.~the baryon density for 
beta-stable matter in the BHF approach (solid line).
The dotted (dashed) lines represent 
calculations for $u$,$d$,$s$ quark matter obtained within the MIT bag model, 
with $B$ parametrized as a Gaussian-like (Woods-Saxon-like) function.
The left, central and right panels correspond to different transition
energy densities $\epsilon_Q = {\rm 0.8,~ 1.1,~ and~ 1.5~ GeV~ fm^{-3}}$.
The full squares (diamonds) are the crossings between 
the hadron and the quark phase.
Two sets of values of the $s$-quark mass and QCD coupling constant 
are considered; $m_s=150~ {\rm MeV}, \alpha_s=0$ (upper panels) and
$m_s=200~ {\rm MeV}, \alpha_s=0.1$ (lower panels). See text for details.
}\label{f:f7}
\end{figure}

\begin{figure}
\includegraphics[width=13cm]{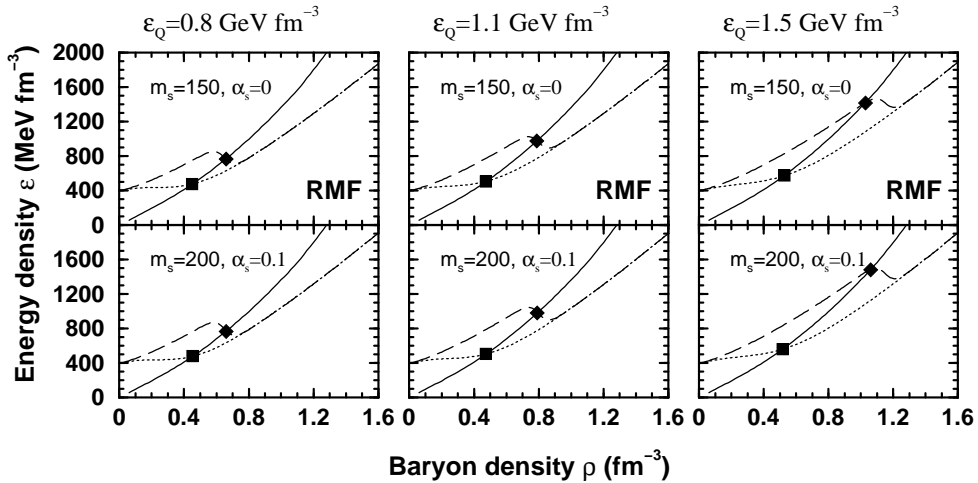}
\caption{Same as Figure 7, but for the RMF baryonic equation of state.
}\label{f:f8}
\end{figure}

\begin{figure}
\includegraphics[width=13cm]{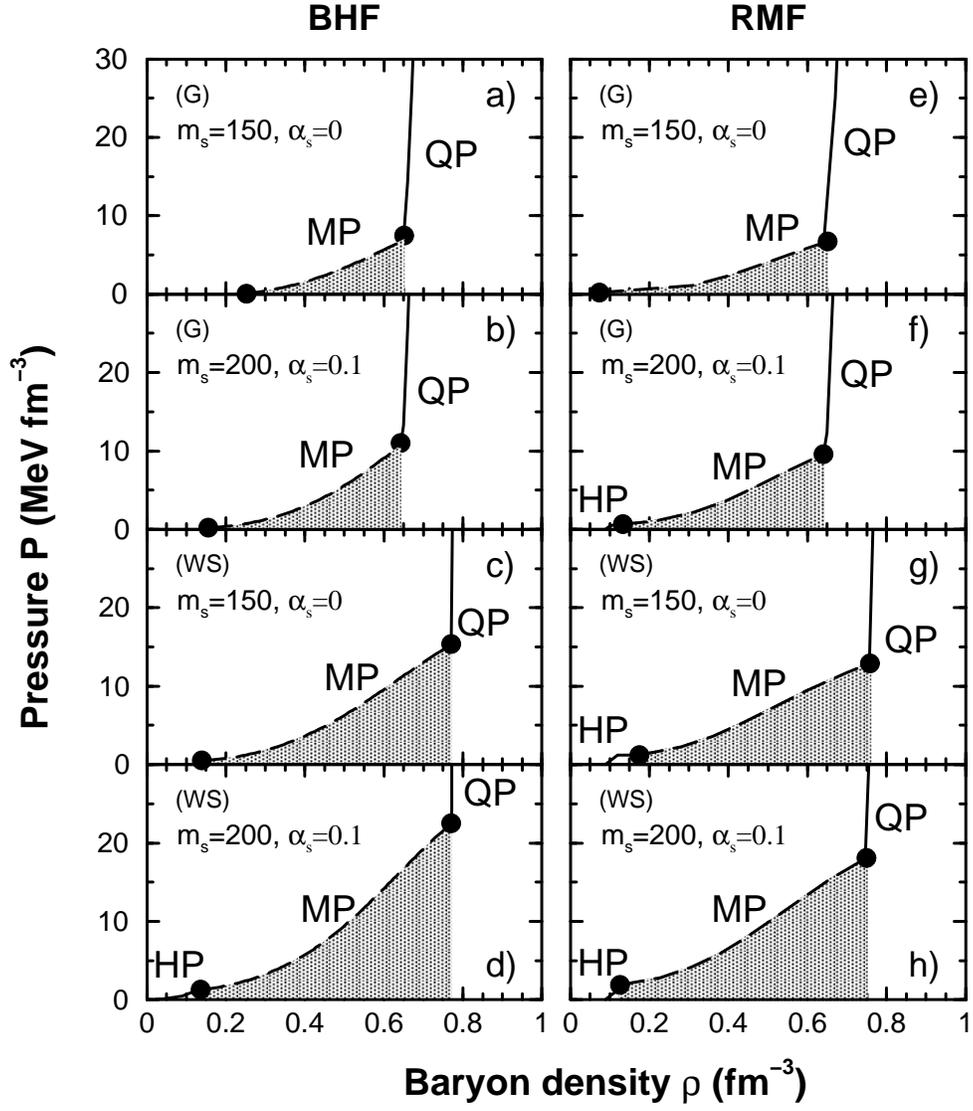}
\caption{The total EOS including both hadronic and quark components is 
displayed for a  transition energy
density $\epsilon_Q = 0.8~ {\rm  GeV ~fm^{-3}}$. 
Different prescriptions for the quark phase are considered,
whereas the hadronic phase is described within the BHF (left-hand panels)
and the RMF (right-hand panels) approaches. In all cases the shaded region,
bordered by two dots, 
indicates the mixed phase MP, while HP and QP label the portions of the EOS 
where pure hadron or pure quark phases are present.
}\label{f:f9}
\end{figure}

\begin{figure}
\includegraphics[width=13cm]{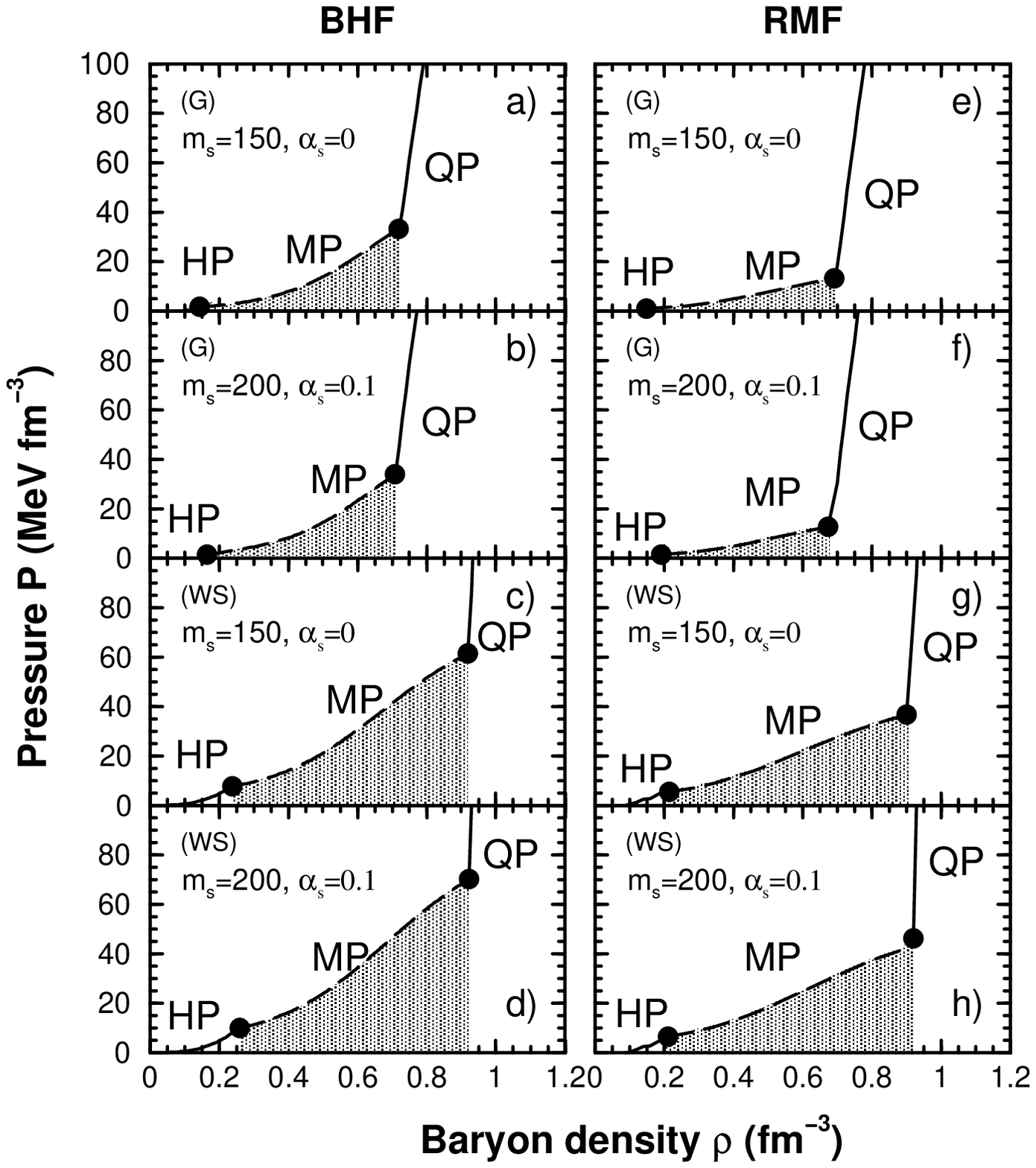}
\caption{Same as figure 9, but for $\epsilon_Q = 1.1~ {\rm GeV~ fm^{-3}}$.
}\label{f:f10}
\end{figure}

\begin{figure}
\includegraphics[width=13cm]{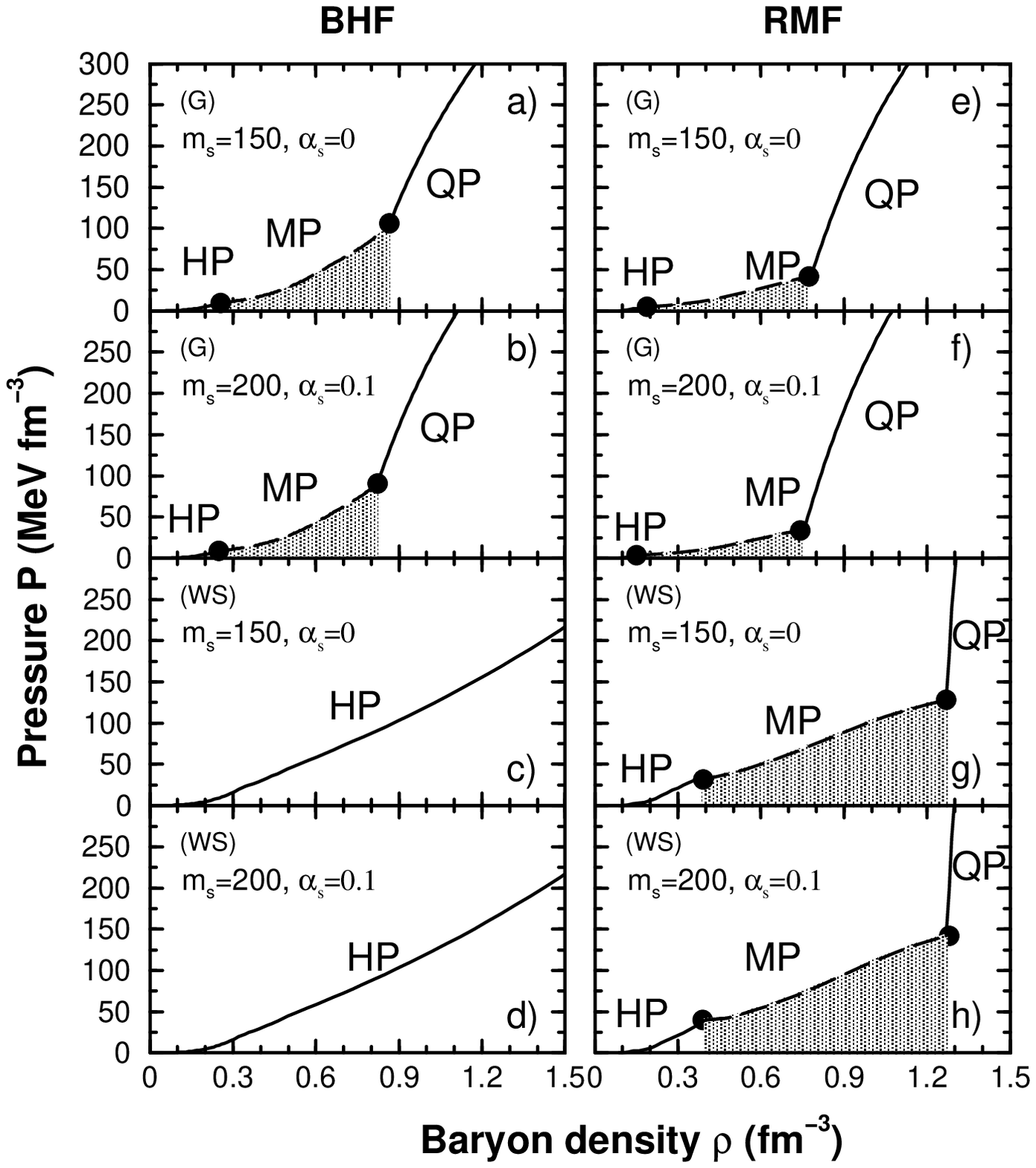}
\caption{Same as figure 9, but for $\epsilon_Q = 1.5~ {\rm GeV~ fm^{-3}}$.
}\label{f:f11}
\end{figure}

\begin{figure}
\includegraphics[width=13cm]{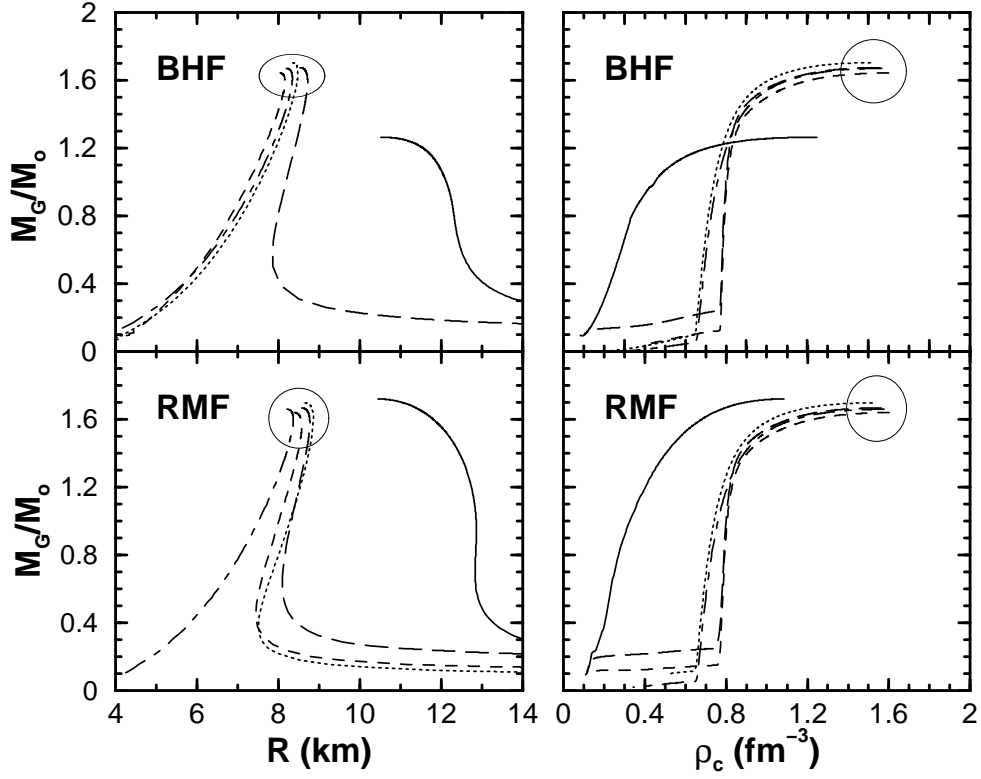}
\caption{The mass-radius (left panels)
and mass-central density (right panels) relations
are displayed for $\epsilon_Q = 0.8~ {\rm GeV~ fm^{-3}}$
and several parametrizations of the bag constant $B$.
In particular the dot-dashed (dotted) lines represent the calculations
performed with $m_s=150~ {\rm MeV}, ~\alpha_s=0$ ($m_s=200~ {\rm MeV},
~\alpha_s=0.1$) and the gaussian-like
parametrization of $B$, whereas the dashed (long-dashed) lines represent the 
calculations performed with $m_s=150~ {\rm MeV}, ~\alpha_s=0$ 
($m_s=200~ {\rm MeV}, ~\alpha_s=0.1$) 
and the Woods-Saxon-like parametrization of $B$.  
Calculations performed with the BHF (RMF) EOS for the hadronic 
component are displayed in the upper (lower) panels by the solid lines.   
}\label{f:f12}
\end{figure}

\begin{figure}
\includegraphics[width=13cm]{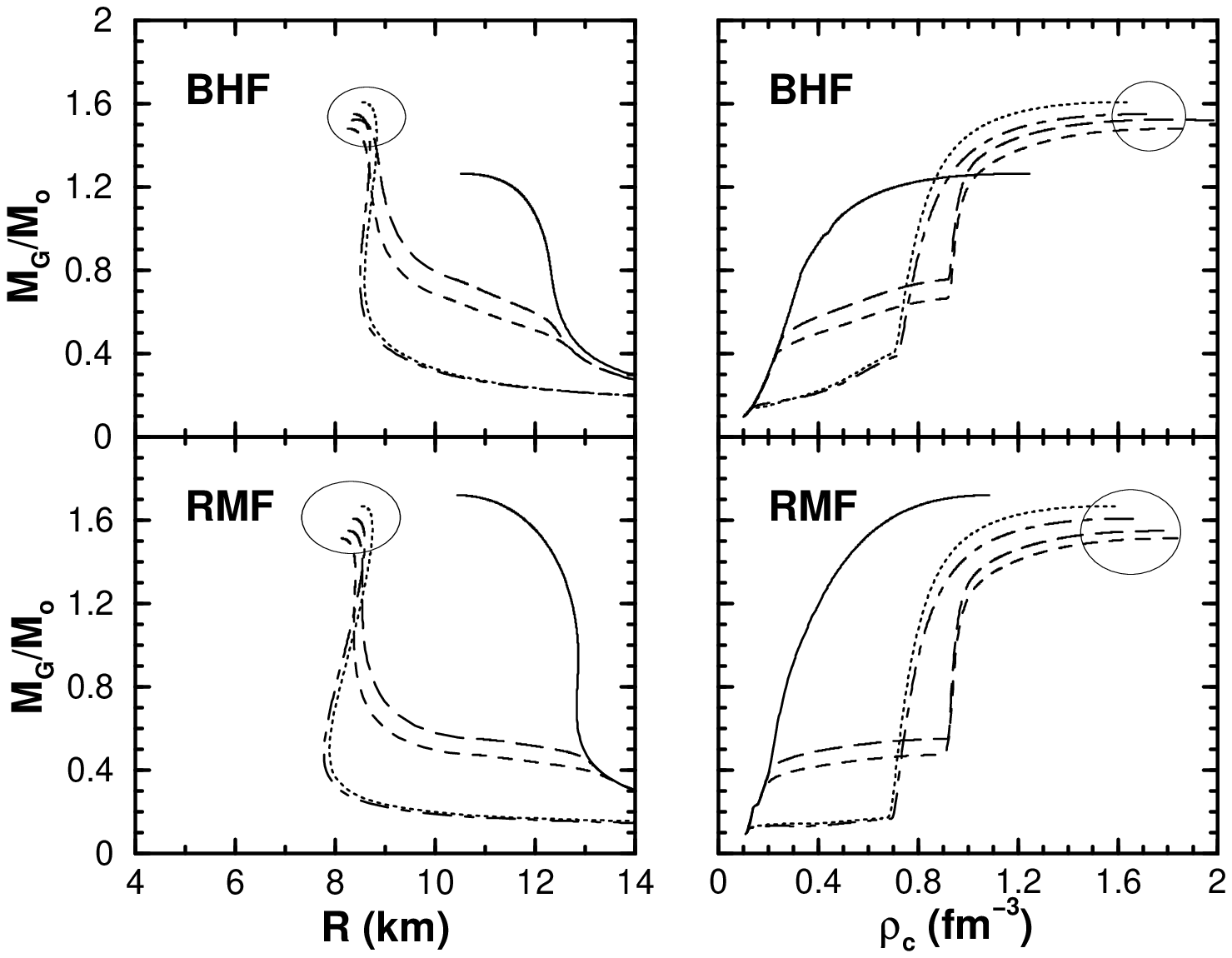}
\caption{Same as figure 12, but for $\epsilon_Q = 1.1~ {\rm GeV~ fm^{-3}}$.
}\label{f:f13}
\end{figure}

\begin{figure}
\includegraphics[width=13cm]{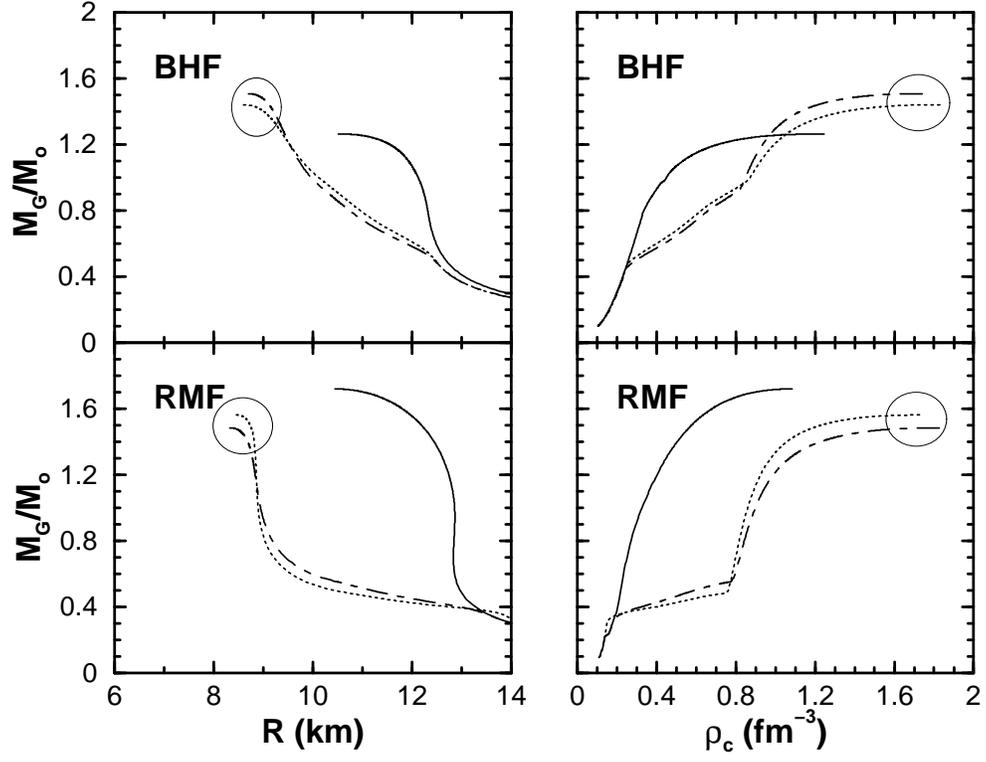}
\caption{Same as figure 12, but for $\epsilon_Q = 1.5~ {\rm GeV~ fm^{-3}}$.
Only the Gaussian-like parametrizations of $B$ are displayed.
}\label{f:f14}
\end{figure}

\begin{figure}
\includegraphics[width=13cm]{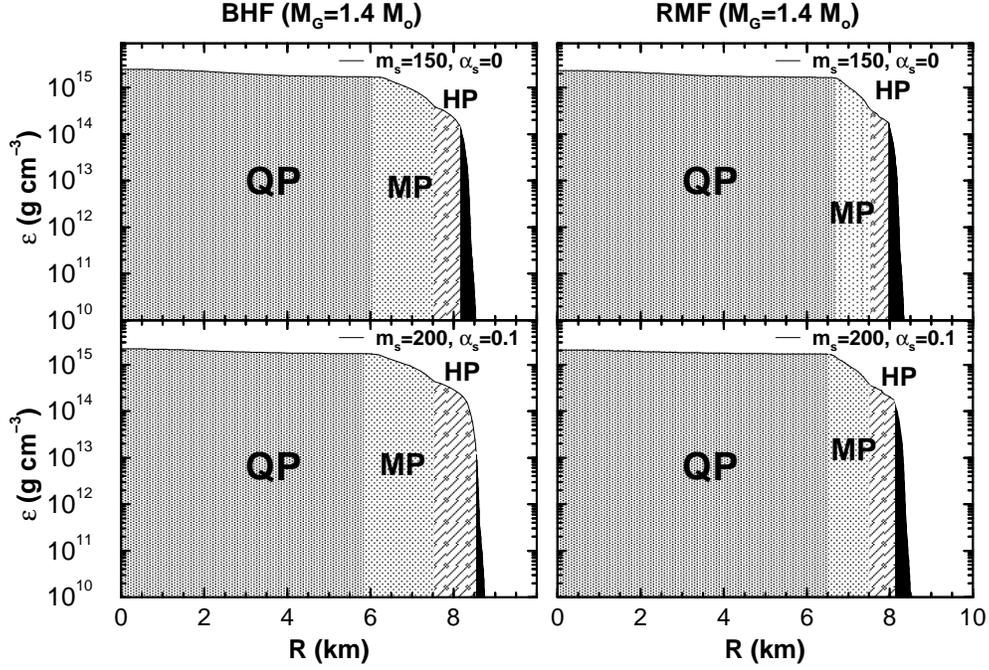}
\caption{Energy density profile of a neutron star with gravitational mass 
$M_G = 1.4~ M_\odot$, obtained for $\epsilon_Q = 1.1~ {\rm GeV~fm^{-3}}$. 
In the left (right) panels the BHF (RMF) EOS has been used for describing 
the hadronic component, and the MIT bag model with a Woods-Saxon 
parametrization of $B$ for the quark phase. 
Upper (lower) panels show the calculations for 
$m_s=150~ {\rm MeV}, \alpha_s=0$ ($m_s=200~ {\rm MeV}, \alpha_s=0.1$).
}\label{f:f15}
\end{figure}

\begin{figure}
\includegraphics[width=13cm]{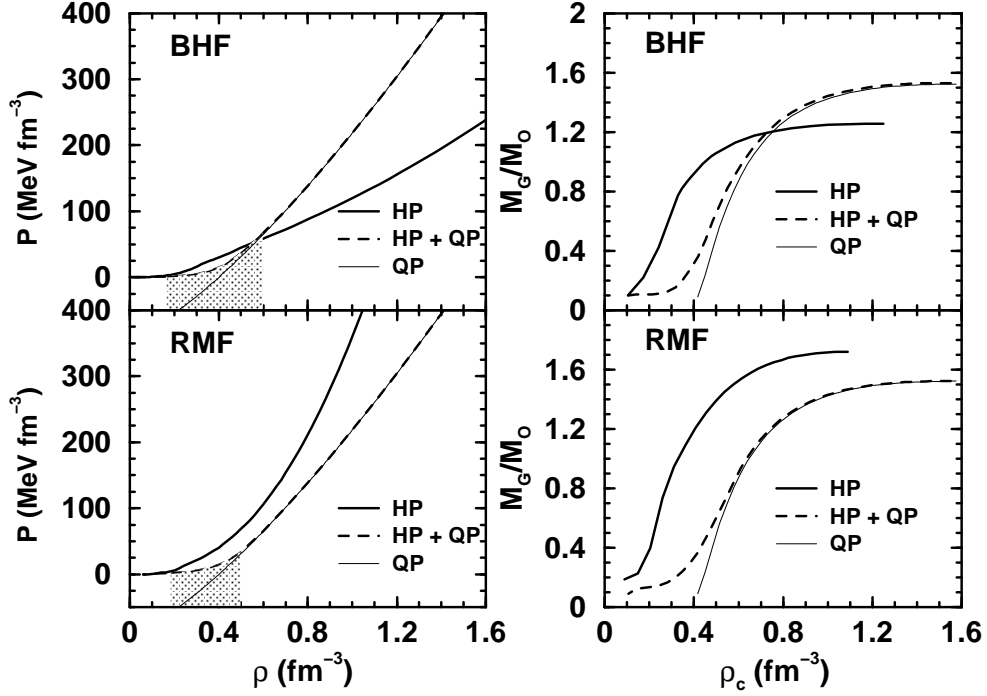}
\caption{In the left panel is shown the EOS for neutron star matter 
(dashed lines labeled by HP+QP) for a density independent value of the 
bag constant $B = {\rm 90~MeV~fm^{-3}}$, with BHF and RMF hadronic equations 
of state. The shaded areas indicate the mixed phase region. 
The corresponding masses vs.~central densities are shown on the right
panels. In all cases the thin and thick lines correspond to the results
obtained for a pure quark and a pure hadron EOS, respectively.
}\label{f:f16}
\end{figure}

\end{document}